\documentclass[%
 reprint,
superscriptaddress,
 amsmath,amssymb,amsfonts,
prl,
twocolumn
]{revtex4-2}

\pdfoutput=1 
\usepackage{graphicx}
\usepackage{bm} 
\usepackage{physics}
\usepackage{xcolor}
\usepackage[colorlinks=true, allcolors=black]{hyperref}

\newcommand{\G}{\mathbb{G}}
\newcommand{\Proj}{\mathbb{P}}

\newcommand{\I}{\mathbb{I}}

\newcommand{\REVISION}[1]{{\color{magenta} #1}}
\DeclareUnicodeCharacter{202F}{\REVISION{There is a 202F unicode character here}}

\begin{document}

\preprint{APS/123-QED}

\title{Can photonic heterostructures provably outperform single-material geometries?}

\author{Alessio~Amaolo}
\affiliation{Department of Chemistry, Princeton University, Princeton, New Jersey 08544, USA}
\email{alessioamaolo@princeton.edu}
\author{Pengning~Chao}
\affiliation{Department of Electrical and Computer Engineering, Princeton University, Princeton, New Jersey 08544, USA}
\author{Thomas~J.~Maldonado}
\affiliation{Department of Electrical and Computer Engineering, Princeton University, Princeton, New Jersey 08544, USA}
\author{Sean~Molesky}
\affiliation{Department of Engineering Physics, Polytechnique Montréal, Montréal, Québec H3T 1J4, Canada}
\author{Alejandro~W.~Rodriguez}
\affiliation{Department of Electrical and Computer Engineering, Princeton University, Princeton, New Jersey 08544, USA}
\date{\today}

\begin{abstract}
Recent advances in photonic optimization have enabled calculation of performance bounds for a wide range of electromagnetic objectives, albeit restricted to single-material systems. 
Motivated by growing theoretical interest and fabrication advances, we present a framework to bound the performance of photonic heterostructures and apply it to investigate maximum absorption characteristics of multilayer films and compact, free-form multi-material scatterers. 
Limits predict trends seen in topology-optimized geometries---often coming within factors of two of specific designs---and may be exploited in conjunction with inverse designs to predict when heterostructures are expected to outperform their optimal single-material counterparts.
\end{abstract}

\maketitle

Large-scale optimization or ``inverse design'' in electromagnetism
involves maximizing a desired field objective (e.g. absorbance,
overlap with a known mode, or scattered power) over thousands to
millions of structural degrees of freedom. The approach has begun to play an important role in the design of high-performing optical
devices~\cite{Molesky_Lin_Piggott_Jin_Vucković_Rodriguez_2018}, leading
to advances in nonlinear frequency conversion~\cite{Sitawarin_Jin_Lin_Rodriguez_2018}, multiplexing~\cite{Piggott_Lu_Lagoudakis_Petykiewicz_Babinec_Vučković_2015},
and bandgap engineering~\cite{Kao_Osher_Yablonovitch_2005}. While
structural optimization is generally NP-hard~\cite{Angeris_Vučković_Boyd_2019}, effectively forbidding guarantees of optimal solutions~\cite{Molesky_Lin_Piggott_Jin_Vucković_Rodriguez_2018}, a new, complementary, approach based on convex relaxations has been shown to provide predictive performance
bounds in a variety of settings~\cite{Angeris_Vučković_Boyd_2019,
  Molesky_Chao_Mohajan_Reinhart_Chi_Rodriguez_2022,
  Chao_Strekha_Kuate_Defo_Molesky_Rodriguez_2022, Kuang_Miller_2020,Gustafsson_Schab_Jelinek_Capek_2020}. 
Examples of this method include recent predictions of maximum angle-integrated absorption~\cite{Molesky_Jin_Venkataram_Rodriguez_2019}, scattering
cross-sections~\cite{Molesky_Chao_Jin_Rodriguez_2020}, communication limits~\cite{miller_optics}, field screening~\cite{Molesky_Chao_Mohajan_Reinhart_Chi_Rodriguez_2022}, dipole
masking~\cite{Molesky_Chao_Mohajan_Reinhart_Chi_Rodriguez_2022}, and local density of states~\cite{Chao_Defo_Molesky_Rodriguez_2023}, among other canonical electromagnetic objectives~\cite{Chao_Strekha_Kuate_Defo_Molesky_Rodriguez_2022}. In addition to
partially assessing the optimality of inverse designs, performance bounds can also yield insights into the physical processes that underpin
desired wave behaviors. For instance, a bound that increases rapidly and then saturates beyond a certain characteristic length suggests that a minimum device size is needed to achieve a given phenomenon. 
Limit information can also be used to guide structural optimization toward optimal
solutions~\cite{Gertler_Kuang_Christie_Miller_2023}. 

The relaxations which allow such bounds to be structure agnostic have limited previous work to a single design material against a background (typically vacuum)~\cite{Chao_Strekha_Kuate_Defo_Molesky_Rodriguez_2022}, precluding investigations of
heterostructures comprising two or more design materials. 
Burgeoning use of these devices for ultrabroadband absorption~\cite{Yang_Ji_Shen_Lee_Zhang_Liu_Guo_2016}, passive cooling~\cite{Raman_Anoma_Zhu_Rephaeli_Fan_2014}, ultrafast photonics~\cite{He_Wang_Zhou_Zhao_Tao_Zhang_2020, Lyu_An_Lin_Qiu_Wang_Chao_Fu_2023}, among other applications~\cite{Soref_2014}, and corresponding efforts to address fabrication challenges, highlight a growing need for definitive statements about the possible advantages of heterostructures. Potentially loose bounds and the NP-hardness of the inverse design problem mean that neither method on its own can prove heterostructure superiority. 
However, the two methods can be combined to provide performance certificates and therefore a new way of assessing the trade-offs associated with multi-material engineering. 

In this article, we show that the duality bounds formalism of
Ref.~\cite{Chao_Strekha_Kuate_Defo_Molesky_Rodriguez_2022} can be extended to handle an arbitrary number of design materials. We present two basic examples that showcase this theory: bounds on the absorbed power from a
plane-wave incident on any multilayer film or from an oscillating dipole in the vicinity of any free-form structure restricted to a square design region. 
Comparisons to topology-optimized designs show heterostructure bounds consistently coming within a factor of two of device performance. More fundamentally, we demonstrate heterostructure designs exhibiting greater performance than either of their corresponding single-material bounds, providing definitive proof that at least in these settings use of multiple materials is advantageous. 

\textit{Formulation:} The key idea underpinning recent bound
optimizations, detailed in Refs.~\cite{Molesky_Jin_Venkataram_Rodriguez_2019,Chao_Strekha_Kuate_Defo_Molesky_Rodriguez_2022,Kuang_Miller_2020,Gustafsson_Schab_Jelinek_Capek_2020}, involves relaxing structural and
physical information in the typical field optimization problem 
by reducing the vector field constraints stated by Maxwell's equations to a user-defined set of scalar constraints ensuring the conservation of power over sub-volumes of the total design region---generalizations of Poynting's theorem~\cite{Chao_Strekha_Kuate_Defo_Molesky_Rodriguez_2022, Chao_Defo_Molesky_Rodriguez_2023}. Convex relaxations and solutions of the resulting quadratically
constrained quadratic program via duality~\cite{Angeris_Vučković_Boyd_2019} or semi-definite
programming~\cite{Miller_2023, Luo_Ma_So_Ye_Zhang_2010} provide a bound on the original
(primal) objective. More precisely, optimization of a given quadratic field objective \(f_0\) over the possible polarization fields \(\ket{\psi_k}\) that may
arise from a set of harmonic fields \(\ket{S_k}\) incident on a given
design region \(V\) can be framed as a quadratic program of the form
\begin{equation}
  \label{eq:opt0}
    \begin{aligned}
        \underset{\{\psi_k\}}{\text{max}} \quad & f_0(\{\ket{\psi_k}\}) \\
        \text{s.t.} \quad & \bra{S_k} \Proj_j \ket{\psi_{k'}} - \bra{\psi_k} \left( \chi_k^{-1 \dagger } \I - \G_0^{(k)\dagger} \right) \Proj_j \ket{\psi_{k'}} = 0, \\
        & \mkern 250mu \forall j,k,k'
    \end{aligned}
\end{equation}
where \(\chi_k\) denotes the electric susceptibility of the design
material at frequency \(\omega_k\), and \(\G_0^{(k)}\)
the corresponding vacuum propagator acting on sources to yield their
respective fields in vacuum---namely, via convolution of the vacuum
Green's function \(G^{(k)}_0(\vb r, \vb r', \omega_k)\) satisfying
\(\dfrac{c^2}{\omega_k^2} \nabla \times \nabla \times G^{(k)}_0(\vb r, \vb r', \omega_k) - G^{(k)}_0(\vb r, \vb r', \omega_k) =
\delta(\vb r - \vb r')\).  Here, \(\I\) and \(\Proj_j\) represent spatial
projections into either the full or a subset \(V_j \in V\) of the design
region \(V\), respectively, and bra-ket notation is used to express complex vector fields over \(V\). The integral constraints in
Eq.~\eqref{eq:opt0} enforce Poynting's theorem---power
conservation---over each selected region. The Lagrange dual of Eq.~\eqref{eq:opt0}
bounds the primal objective~\cite{Boyd_Vandenberghe_2004}.

Extending this formalism to multiple materials may be carried out as follows. Expressing \(\ket{\psi_k}=\sum_m \ket{\psi_{k,m}}\) as a sum
of polarization currents associated with a given material \(m\) of
susceptibility \(\chi_{k,m}\), it suffices to allow each region to be filled with any
combination of the selected \(\chi_{k,m}\). To avoid unphysical solutions,
however, one must also enforce an additional constraint precluding the overlap of distinct materials: the polarization
currents associated with distinct materials must be orthogonal in any sub-region. 
Performing this modification, the optimization problem becomes:
\begin{equation}
\label{eq:opt}
\begin{aligned}
  \underset{\{\psi_{k,m}\}}{\text{max}} \qquad & f_0(\{\ket{\psi_{k,m}}\}) \\
    \text{s.t.} \qquad & \sum_m \Big( \bra{S_k} \Proj_j \ket{\psi_{k',m}} - \bra{\psi_{k,m}} \Proj_j \chi_{k,m}^\dagger \ket{\psi_{k',m}} \\
    \qquad &- \sum_{m'} \bra{\psi_{k,m}} - \Proj_j \G_0^{(k)\dagger} \ket{\psi_{k',m'}} \Big) = 0 \quad \forall j,k,k', \\ 
    & \bra{\psi_{k,m}} \Proj_j \ket {\psi_{k',m'}} = 0 \quad \forall j, k, k', m\neq m'.
\end{aligned}
\end{equation}

\begin{figure}
    \centering
    \includegraphics[width=0.48\textwidth]{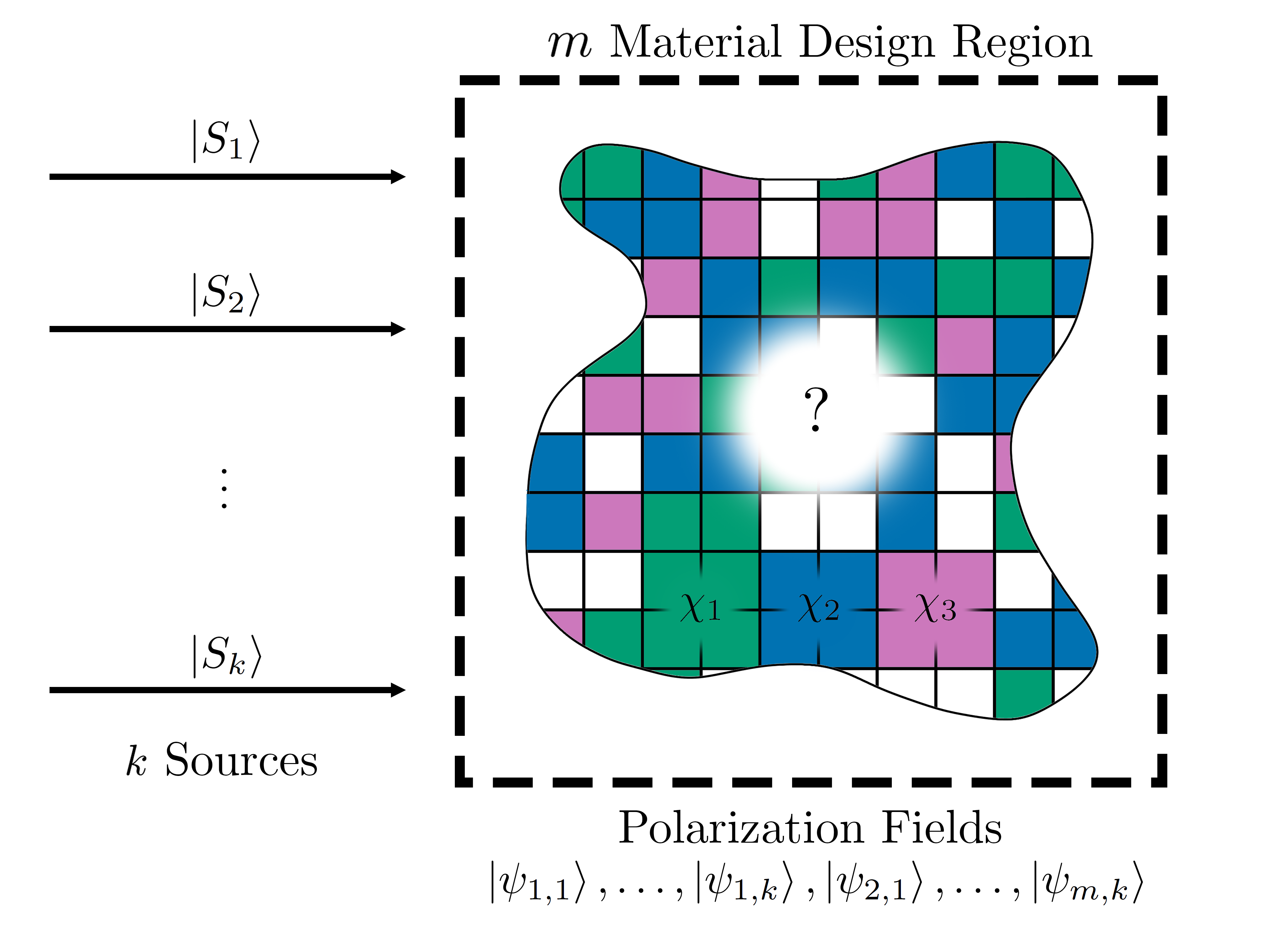}
    \caption{$k$ sources are incident on an $m$ material design region, and polarization currents break down into $k \times m$ components. Each computational voxel in the design region can be any of $m$ materials (exclusively) or background (vacuum). Objectives are written in terms of polarization currents.}
    \label{fig:cartoon}
\end{figure}

A schematic of the problem under consideration is shown in Fig.~\ref{fig:cartoon}. Detailed calculations of the corresponding Lagrange dual and gradients, as well as a proof of the existence of dual solutions are provided in the Supplemental Material (SM) \cite{supp}. 

The remainder of the article showcases the utility of Eq.~\eqref{eq:opt} by computing bounds
on two representative scattering problems: maximization of absorption of a planewave or a dipolar source. 
Resulting bounds are compared to the best performing devices obtained via topology optimization, as
outlined in~\cite{Christiansen_Sigmund_2021} for single materials and in the SM~\cite{supp} for multiple materials. 
In all cases, structural optimization routines and bounds calculations were initialized with an empty design region or as described in the SM~\cite{supp}, respectively. 

\begin{figure*}
  \includegraphics[width=\textwidth]{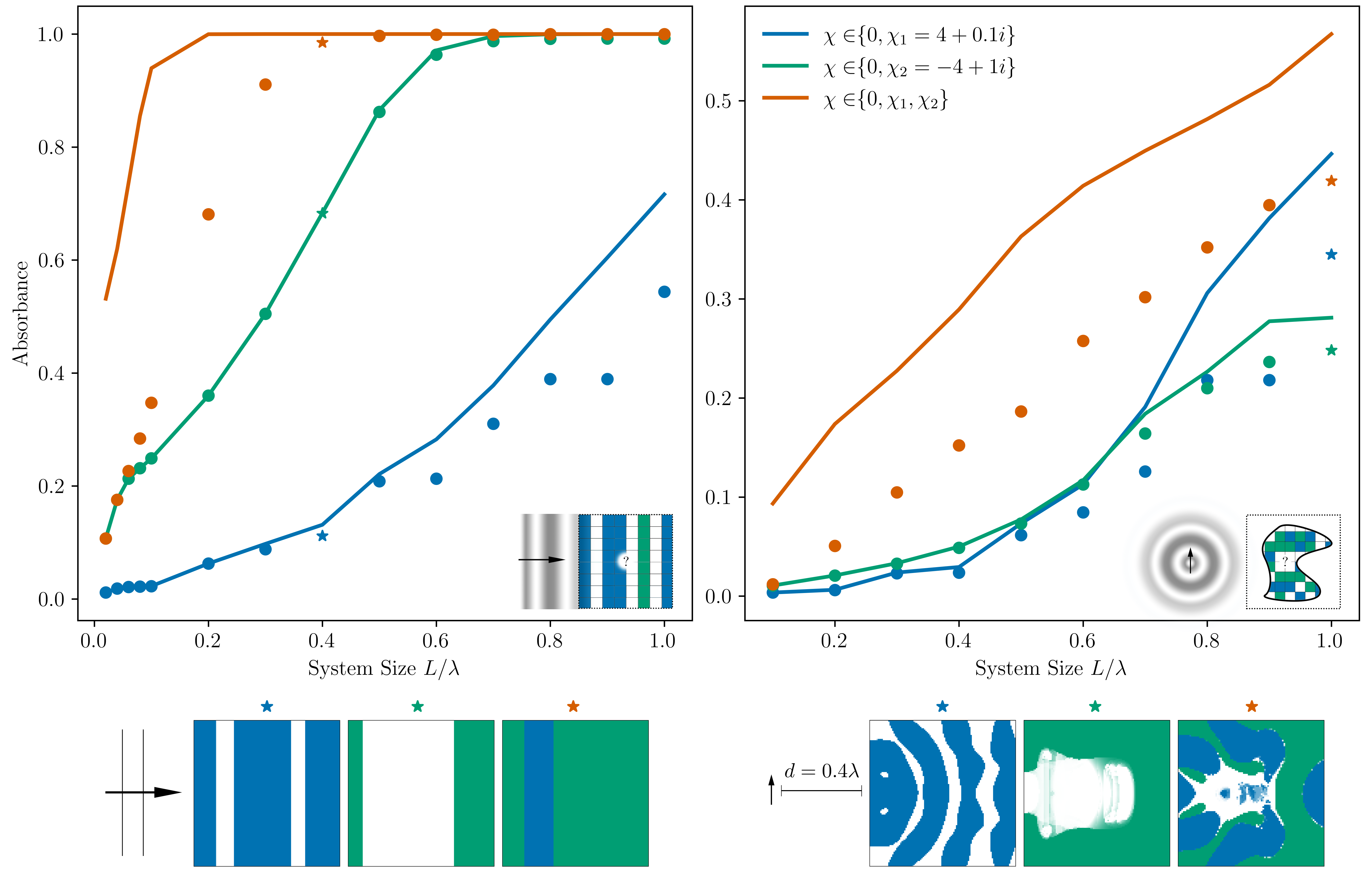}
  \caption{Bounds on maximum absorption (solid lines) and associated inverse designs (dots) for either a planewave incident on multilayer films (left) or a 
  an oscillating dipole near structures contained within a square design region (right). Absorbance values quantify power absorbed in the respective design regions normalized to the incident power for the planewave or emitted power of the dipole in vacuum, for either single-material or heterogeneous systems composed of \(\chi_1 = 4+0.1i\) (dielectric), \(\chi_2 = -4+1i\) (metal), and vacuum regions. Representative inverse designs (\(L=0.4\lambda\) for films and \(L=1\lambda\) for compact structures) are labeled with stars and shown below the main plots. The colors representing each material are chosen to be the same for both inverse designs and bounds. Additional optimized structures are shown in the SM~\cite{supp}.}
  \label{fig:abs_combined}
\end{figure*}

\textit{Planewave incident on a multilayer film:} We first consider a
TM polarized planewave with wavelength \(\lambda\) incident on a device
of length \(L\leq \lambda\) consisting of multiple layers of variable
thicknesses. We seek to maximize the ratio of absorbed to incident
power. The computational ease of this effectively one dimensional problem allows us
to readily enforce the constraints in Eq.~\eqref{eq:opt} over each pixel
of the computational grid. We consider use of either a
single dielectric of \(\chi = 4+0.1i\), a single metal of \(\chi=-4+1i\), or the use of both materials.

Results are shown in Fig.~\ref{fig:abs_combined}~(left) and compared to
topology-optimized inverse designs. Notably, performance values for
optimized structures and associated bounds reflect the increased
ability of thicker films and greater material choice to achieve
higher planewave absorption, which eventually saturates to 100\% for
sufficiently large devices. We note that the designs of size \(L/\lambda = 0.1\) were verified to be globally optimal by brute force, and that all bounds approach zero as \( L/\lambda \to 0 \).
Heterostructure bounds are found to correctly anticipate the earlier onset of
perfect absorbance and devices' provably superior performance compared to their single-material counterparts.
This increased performance is evident at
intermediate thicknesses \(0.1\lambda \leq L \leq 0.5\lambda\) but
rapidly vanishes in highly subwavelength or optically thick films,
with metallic films outperforming their dielectric counterparts, and
heterostructures primarily if not entirely composed of 
metal. As expected, metallic devices outperform dielectrics at small \(L/\lambda\) by exploiting plasmonic confinement, with dielectrics becoming increasingly effective as \(L \to \lambda\).
Optimized designs suggest that the primary mechanism by which hybrid heterostructures outperform single-material structures at intermediate \(L/\lambda\) is by the creation of metallic cavities filled with dielectric (as opposed to vacuum) absorbing material. 
As seen from the optimized
multilayer structure of thickness \(L=0.4\lambda\) shown on the inset
(orange star), there appears to be little to no use of vacuum regions
within the design domain, indicating that optimal performance may be
predicted from a single-material framework, with either the metal or
dielectric as the background medium. 
More examples of this phenomenon, which is likely specific to the case of maximizing absorption, are presented in the SM~\cite{supp}. This could in theory be studied by modifying the single-material bounds formalism to take into account a design region surrounded by vacuum but comprising exclusively the design materials (generalizing ideas in Ref.~\cite{Gertler_Kuang_Christie_Miller_2023}).  
Lastly, we note the non-additivity of the bounds: calculated limits on heterostructure absorbance are not given by the sum of the single material limits. 
In fact, topology optimized inverse designs outperform the sum of the two single-material bounds when \( 0.08\lambda \leq L \leq 0.5\).

\textit{Dipole near a compact structure:} Next, we consider a TM
polarized dipole oscillating at frequency \( \lambda/c \) and located a
distance \( d=0.4\lambda \) from a square design region of side length \( L
\leq \lambda \). We now seek to maximize the ratio of absorbed power to
the power radiated by the dipole in vacuum. For computational
convenience, the energy conservation constraints in Eq.~\eqref{eq:opt} are only
enforced in \(100\) equally sized sub-domains for single materials and \(25\) sub-domains for two materials. Again, we consider the use
of a dielectric with \(\chi = 4+0.1i\), a metal with \(\chi=-4+1i\), and
the use of both materials.

Results are shown in Fig.~\ref{fig:abs_combined}~(right) and compared to
topology-optimized inverse designs. Similar to the multilayer
examples, bounds are seen to accurately predict trends in device
performance, with the non-additive nature of multiple-material bounds
becoming particularly pronounced for wavelength-scale devices. 
The performance of two-material devices is provably better than any single-material structure for \(0.1 \lambda \leq L \leq 0.9 \lambda\). 
Inspecting bound solutions and inverse designs, we postulate that the dielectric (blue star) and metallic device (green star) of size \( L/\lambda = 1 \) work by reflecting incident fields into an absorbing block near the dipole and by forming a cavity mode, respectively.  
As seen from the representative
optimized structure shown on the inset (orange star), we find that
greater performance in this setting is achieved through use of vacuum,
dielectric, and metallic regions combined. 
The structure appears to be a hybrid of the two single-material structures, highlighting that optimal structures may be combinations of single-material structures that leverage the benefits of both materials. Interestingly, this is the only multi-material structure that significantly utilizes vacuum, as well as the only \( L/\lambda \) where the dielectric single-material device greatly outperforms the corresponding metallic device. Rather than replacing vacuum in a metallic cavity with absorbing dielectric (as was the case for multilayer films or for structures with \( L/\lambda < 1 \) in this setting), the primary absorbing mechanism here appears to be dielectric confinement.

\textit{Concluding remarks:} This work shows that our extension to the duality bounds framework~\cite{Chao_Strekha_Kuate_Defo_Molesky_Rodriguez_2022} for handling multi-material settings is both effective and useful. 
For the studied applications, access to multiple materials is seen to provide non-trivial performance benefits over single-material systems, except in very large designs where increased structural freedom makes greater material choice less consequential. 
While there is no proof of strong duality, the observation of optimized structures exceeding single-material bounds provides evidence that structures composed of multiple media offer meaningful advantages for photonic design. 
Examples where inverse designs converge on structures with three different refractive indices further motivates use of multi-material physical limits. 
Promising applications of this theory include problems involving multiple sources that may be separately addressed by distinct material responses (e.g., with multiple dispersive materials).

We acknowledge the support by the National Science Foundation under the Emerging Frontiers in Research and Innovation (EFRI) program, Award No. EFMA-164098, the Defense Advanced Research Projects Agency (DARPA) under Agreements No. HR00111820046, No. HR00112090011, and No. HR0011047197, and by a Princeton SEAS Innovation Grant. 
SM also acknowledges financial support from IVADO (Institut de valorisation des données, Québec).
The simulations presented in this article were performed on computational resources managed and supported by Princeton Research Computing, a consortium of groups including the Princeton Institute for Computational Science and Engineering (PICSciE) and the Office of Information Technology's High Performance Computing Center and Visualization Laboratory at Princeton University.
The views, opinions and findings expressed herein are those of the authors, and should not be interpreted as representing the official views or policies of any institution.

\bibliography{citations}
\end{document}


\title{Supplemental Material for ``Can photonic heterostructures provably outperform single-material geometries?"}

\author{Alessio Amaolo}
\affiliation{Department of Chemistry, Princeton University, Princeton, New Jersey 08544, USA}
\author{Pengning Chao}
\affiliation{Department of Electrical and Computer Engineering, Princeton University, Princeton, New Jersey 08544, USA}
\author{Thomas J. Maldonado}
\affiliation{Department of Electrical and Computer Engineering, Princeton University, Princeton, New Jersey 08544, USA}
\author{Sean Molesky}
\affiliation{Department of Engineering Physics, Polytechnique Montréal, Montréal, Québec H3T 1J4, Canada}
\author{Alejandro W. Rodriguez}
\affiliation{Department of Electrical and Computer Engineering, Princeton University, Princeton, New Jersey 08544, USA}
\date{\today}

\preprint{APS/123-QED}

\maketitle
\section{Computing the Lagrange Dual}
The primal optimization problem takes the form
\begin{equation}
\label{eq:optproblem}
\begin{aligned}
  \underset{\{T{k,m}\}}{\text{max}} \qquad & f_0(\{\ket{T_{k,m}}\}) \\
    \text{s.t.} \qquad & \sum_m \Big( \bra{S_k} \Proj_j \ket{T_{k',m}} - \bra{T_{k,m}} \Proj_j \chi_{k,m}^\dagger \ket{T_{k',m}} - \sum_{m'} \bra{T_{k,m}} - \Proj_j \G_0^{(k)\dagger} \ket{T_{k',m'}} \Big) = 0 \quad \forall j,k,k', \\ 
    & \bra{T_{k,m}} \Proj_j \ket {T_{k',m'}} = 0 \quad \forall j, k, k', m\neq m'.
\end{aligned}
\end{equation}
This notation differs from the main text via \(\psi_{k,m} \to T_{k,m}\). As in the main text, \(\ket{S_k}\) is a source \(k\), the polarization current due to source \(k\) is \(\ket{T_k} = \sum_m \ket{T_{k,m}}\) with \(\ket{T_{k,m}}\) the polarization current due to source \(k\) and material \(m\) and is defined in the design region \(V\). \(\chi_{k,m}\) is the susceptibility of material \(m\) at \(\omega_k\), and \(\G^{(k)}_0\) is
the corresponding vacuum propagator acting on sources to yield their
corresponding fields in vacuum---namely, via convolution of the vacuum
Green's function \(G^{(k)}_0(\vb r, \vb r', \omega_k)\) satisfying
\(\dfrac{c^2}{\omega_k^2} \nabla \times \nabla \times G^{(k)}_0(\vb r, \vb r', \omega_k) - G^{(k)}_0(\vb r, \vb r', \omega_k) =
\delta(\vb r - \vb r')\). \(\I\) and \(\Proj_j\) represent spatial
projections onto either the full or a subset \(V_j \in V\) of the design
region \(V\), respectively. Lastly, \(f_0\) is a quadratic function of the polarization currents \(\ket{T_{k,m}}\). 

For the first constraint, we will take the real and imaginary parts and write the Lagrange multiplier corresponding to a given \(j,k,k'\) as \(\lambda_{R/I}^{j,k,k'}\) (symmetric and asymmetric constraints respectively). For the second, we will use Lagrange multipliers \(\lambda_{SO/AO}^{k,k',m,m'}\) (symmetric/asymmetric orthogonal constraint). 
Now we can write 
\begin{equation}
\label{eq:lagrangian}
    \Lag (T,S) = 
    \begin{bmatrix}
    \bra{T_{opt}} & \bra{S} 
    \end{bmatrix}
    \begin{bmatrix}
    -Z^{TT}(\lambda) & Z^{TS}(\lambda) \\
    Z^{ST}(\lambda) & 0  
    \end{bmatrix} 
    \begin{bmatrix}
    \ket{T_{opt}} \\
    \ket{S} 
    \end{bmatrix} ,
\end{equation}
where \(\Lag\) is the Lagrangian, \(\ket{T_{opt}} = \begin{bmatrix} \ket{T_{1,1}} & \ket{T_{1,2}} & \ldots & \ket{T_{1,m}} & \ldots & \ket{T_{2,1}} & \ldots &\ket{T_{n_s, m}} \end{bmatrix}^T\)
for \(n_s\) sources and \(m\) materials, \( \ket{S} = \begin{bmatrix} \ket{S_1} & \ldots & \ket{S_{n_s}} \end{bmatrix}^T \), and \(Z^{TT}\) and \(Z^{TS} = Z^{ST \dagger}\) matrices represent the quadratic and linear parts of the Lagrangian, respectively. 
We also denote \(N\) the numerical length of a single \(\ket{T_{k,m}}\) vector. 

Writing out the constraints we find
\begin{equation}
\begin{aligned} 
    Z^{TS} = \Obj_{lin} + \sum_j \dfrac12 &\begin{bmatrix}
    \begin{bmatrix}
    \lambda_R^{j,1,1} \Proj_j & \ldots & \lambda_R^{j,n_s,1} \Proj_j \\
    \vdots & \vdots & \vdots \\ 
    \lambda_R^{j,1,1} \Proj_j & \ldots & \lambda_R^{j,n_s,1} \Proj_j \\
    \end{bmatrix} \\
    \vdots \\ 
    \begin{bmatrix}
    \lambda_R^{j,1,n_s} \Proj_j & \ldots & \lambda_R^{j,n_s,n_s} \Proj_j \\
    \vdots & \vdots & \vdots \\ 
    \lambda_R^{j,1,n_s} \Proj_j & \ldots & \lambda_R^{j,n_s,n_s} \Proj_j \\
    \end{bmatrix} \\
    \end{bmatrix} \\
    - \sum_j \dfrac{1}{2i} &\begin{bmatrix}
    \begin{bmatrix}
    \lambda_I^{j,1,1} \Proj_j & \ldots & \lambda_I^{j,n_s,1} \Proj_j \\
    \vdots & \vdots & \vdots \\ 
    \lambda_I^{j,1,1} \Proj_j & \ldots & \lambda_I^{j,n_s,1} \Proj_j \\
    \end{bmatrix} \\
    \vdots \\ 
    \begin{bmatrix}
    \lambda_I^{j,1,n_s} \Proj_j & \ldots & \lambda_I^{j,n_s,n_s} \Proj_j \\
    \vdots & \vdots & \vdots \\ 
    \lambda_I^{j,1,n_s} \Proj_j & \ldots & \lambda_I^{j,n_s,n_s} \Proj_j \\
    \end{bmatrix} \\
    \end{bmatrix}, \\
\end{aligned} 
\end{equation} 
\begin{equation}
    Z^{ST} = Z^{TS \dagger},
\end{equation}
where \(\Obj_{lin}\) is the linear part of the objective.

\begin{equation}
\begin{aligned}
    Z^{TT} = \; &\Obj_{quad} \\
    &+ \sum_j \begin{bmatrix} 
    R_{1,1}^j & \ldots & R_{1,n_s}^j \\ 
    \vdots & \ddots & \vdots \\ 
    R_{n_s,1}^j & \dots & R_{n_s, n_s}^j
    \end{bmatrix} + \sum_j \begin{bmatrix} 
    I_{1,1}^j & \ldots & I_{1,n_s}^j \\ 
    \vdots & \ddots & \vdots \\ 
    I_{n_s,1}^j & \dots & I_{n_s, n_s}^j
    \end{bmatrix} + \sum_j \begin{bmatrix} 
    S_{1,1}^j +A_{1,1}^j & \ldots & S_{1,n_s}^j + A_{1,n_s}^j \\ 
    \vdots & \ddots & \vdots \\ 
    S_{n_s,1}^j + A_{n_s,1}^j & \dots & S_{n_s, n_s}^j + A_{n_s, n_s}^j 
    \end{bmatrix},
\end{aligned}
\end{equation}
with 
\begin{subequations}
\begin{align} 
R_{k,k'}^j = R_{k',k}^{j\dagger} = \dfrac12 \lambda_R^{j,k,k'} &\begin{bmatrix} 
\Proj_j (\chi_{k,1}^{-1 \dagger} - \G_0^{(k)\dagger}) & \dots & -\Proj_j \G_0^{(k)\dagger} \\ 
\vdots & \ddots & \vdots \\ 
-\Proj_j \G_0^{(k)\dagger} & \dots & \Proj_j (\chi_{k,m}^{-1 \dagger} - \G_0^{(k)\dagger}) \\ 
\end{bmatrix} \nonumber \\
+ \dfrac12 \lambda_R^{j,k',k} &\begin{bmatrix} 
\Proj_j (\chi_{k',1}^{-1} - \G_0^{(k')}) & \dots & -\Proj_j \G_0^{(k')} \\ 
\vdots & \ddots & \vdots \\ 
-\Proj_j \G_0^{(k')} & \dots & \Proj_j (\chi_{k',m}^{-1} - \G_0^{(k')}) \\
\end{bmatrix}, \\
I_{k,k'}^j = I_{k',k}^{j\dagger} = \dfrac{1}{2i} \lambda_I^{j,k,k'} &\begin{bmatrix} 
\Proj_j (\chi_{k,1}^{-1 \dagger} - \G_0^{(k)\dagger}) & \dots & -\Proj_j \G_0^{(k)\dagger} \\ 
\vdots & \ddots & \vdots \\ 
-\Proj_j \G_0^{(k)\dagger} & \dots & \Proj_j (\chi_{k,m}^{-1 \dagger} - \G_0^{(k)\dagger}) \\ 
\end{bmatrix} \nonumber \\
- \dfrac{1}{2i} \lambda_I^{j,k',k} &\begin{bmatrix} 
\Proj_j (\chi_{k',1}^{-1} - \G_0^{(k')}) & \dots & -\Proj_j \G_0^{(k')} \\ 
\vdots & \ddots & \vdots \\ 
-\Proj_j \G_0^{(k')} & \dots & \Proj_j (\chi_{k',m}^{-1} - \G_0^{(k')}) \\
\end{bmatrix},
\end{align}
\end{subequations}
with  \(-\Proj_j \G_0^{(k)\dagger}\) or \(-\Proj_j \G_0^{(k')}\) present in every off-diagonal element. Furthermore,
\begin{subequations}
\begin{align}
S_{k,k'}^j = \dfrac12 &\begin{bmatrix}
0 & \Proj_j \lambda_{SO}^{j,k,k',1,2} & \dots & \Proj_j \lambda_{SO}^{j,k,k',1,m} \\ 
 \Proj_j \lambda_{SO}^{j,k,k',1,2} & 0 & \dots & \Proj_j \lambda_{SO}^{j,k,k',2,m} \\ 
\vdots & \vdots & \ddots & \vdots \\ 
 \Proj_j \lambda_{AO}^{j,k,k',1,m} & \Proj_j \lambda_{SO}^{j,k',2,m} & \dots & 0
\end{bmatrix} = {S_{k,k'}^j}^T = S_{k',k}^j, \\ 
A_{k,k'}^j = \dfrac{1}{2i} &\begin{bmatrix}
0 & \Proj_j \lambda_{AO}^{j,k,k',1,2} & \dots & \Proj_j \lambda_{AO}^{j,k,k',1,m} \\ 
-\Proj_j \lambda_{AO}^{j,k,k',1,2}  & 0 & \dots & \Proj_j \lambda_{AO}^{j,k,k',2,m} \\ 
\vdots & \vdots & \ddots & \vdots \\ 
-\Proj_j \lambda_{AO}^{j,k,k',1,m} & -\Proj_j \lambda_{AO}^{j,k,k',2,m} & \dots & 0
\end{bmatrix} = {A_{k,k'}^j}^\dagger = -A_{k',k}^j. \\ 
\end{align}
\end{subequations}
\(\Obj_{quad}\) is the quadratic part of the objective and all \(R,I,S,A\) are \(m\times m\) block matrices of \(N \times N\) matrices.
We can compute the dual function \(\mathcal{G}\):
\begin{equation}
    \mathcal{G}(\lambda) = \sup_{\ket{T}} \Lag(\lambda, T).
    \label{def:dual}
\end{equation}
We find the stationary point \(\ket{T^*}\) of \(\Lag\) by solving the relation 
\begin{equation}
    \pdv{\Lag}{\bra{T^*}} = 0,
\end{equation}
which leads to the linear system
\begin{equation}
    Z^{TT}\ket{T^*} = Z^{TS}\ket{S}.
\end{equation}
In order for the dual to be finite, \(Z^{TT}\) must be positive definite, so this linear system is invertible, leading to
\begin{equation}
    \ket{T^*} = Z^{TT-1} Z^{TS}\ket{S},
\end{equation}
\begin{equation}
    \mathcal{G}(\lambda) = \bra{S}Z^{ST} Z^{TT-1} Z^{TS} \ket{S}.
    \label{eq:dual_ZTT}
\end{equation}
Finally,
\begin{equation}    \pdv{\mathcal{G}}{\lambda_i} = 2\Re{\bra{T^*}\pdv{Z^{TS}}{\lambda_i} \ket{S}} - \bra{T^*} \pdv{Z^{TT}}{\lambda_i} \ket{T^*}.
    \label{eq:dual_derivative}
\end{equation}
In the specific case of maximizing absorption, \(\Obj_{lin} = 0\). Absorbed power is 
\begin{equation} 
\sum_{k,m} \dfrac{Zc^2}{2 \omega_k^2} \bra{T_{k,m}} \dfrac{\Im \chi_{k,m}}{\abs{ \chi_{k,m}}^2} \ket{T_{k,m}}, 
\end{equation} 
with \(Z\) the vacuum impedance, giving us \(\Obj_{quad}\) (being careful of the negative sign in Eq.~\eqref{eq:lagrangian}). Each term can be normalized as desired by the incident or total power. 

\section{Solving the Dual Problem}
In order to solve the convex dual problem using an interior point method, we must first find an initial feasible point. 
This requires choosing \(\lambda\) such that \(Z^{TT}\) is positive definite, thereby ensuring that \(Z^{TT}\) is invertible and therefore that the dual is well defined. This can be done reliably by leveraging the fact that the imaginary part of the Maxwell Green's function, \(\Im \G_0^{(k)}\), is positive semidefinite \cite{tsang_scattering_2004}. Take \(\Proj_{j=0} = \I\) to be the projector corresponding to global constraints, which is always enforced in our implementation. By setting all Lagrange multipliers \(\lambda=0\) except \(\lambda_I^{j=0,k,k}\) for all \(k\), we set \(R_{k,k'}^j = S_{k,k'}^j = A_{k,k'}^j = 0\) for all \(k,k',j\). This also sets \(I_{k,k'}^j = 0\) for all \(j, k \neq k'\) and, although not necessary, for all \(j \neq 0, k=k'\).
The quadratic part of \(\Lag\) becomes
\begin{equation}
    Z^{TT} = \Obj_{quad} + \begin{bmatrix}
        I_{1,1}^0 & \ldots & 0 \\ 
        \vdots & \ddots & \vdots \\ 
        0 & \ldots & I_{n_s, n_s}^0
    \end{bmatrix},
\end{equation}
with \(I_{k,k}^0\) taking the following block diagonal form:
\begin{equation}
\begin{aligned}
    I_{k,k}^0 &= \lambda_I^{0,k,k} \begin{bmatrix}
        \Asym \left( \I \chi_{k,1}^{-1 \dagger} - \G_0^{(k)\dagger} \right) & \ldots & 0 \\ 
        \vdots & \ddots & \vdots \\ 
        0 & \ldots & \Asym \left( \I \chi_{k,m}^{-1\dagger} - \G_0^{(k)\dagger} \right) \\
    \end{bmatrix} \\
    &= \lambda_I^{0,k,k} \begin{bmatrix}
        \I \dfrac{\Im \chi_{k,1}}{\abs{\chi_{k,1}}^2} + \Asym \G_0^{(k)} & \ldots & 0 \\ 
        \vdots & \ddots & \vdots \\ 
        0 & \ldots & \I \dfrac{\Im \chi_{k,1}}{\abs{\chi_{k,m}}^2} + \Asym \G_0^{(k)}
    \end{bmatrix},
\end{aligned}
\end{equation}
which is positive definite for \(\Im \chi_{k,m} > 0\). 
Therefore, as long as all materials have some loss (as they do in the main text), we can simply increase \(\lambda_I^{0,k,k}\) for all \(k\) to make \(Z^{TT}(\lambda)\) positive definite for any finite \(\Obj_{quad}\). 
We initialize all bounds calculations in the main text (where \(n_s = 1\)) by setting \(\lambda_I^{0,k,k} = 1 \quad \forall k\). 
When \(\Im \chi_{k,m} = 0\), we may be able to generalize the technique utilized in Ref~\cite{Chao_Defo_Molesky_Rodriguez_2023} (Supporting Information, Section 11). 

\section{Dependence on number of sub-region constraints}
\begin{figure}
    \centering\includegraphics[width=\textwidth]{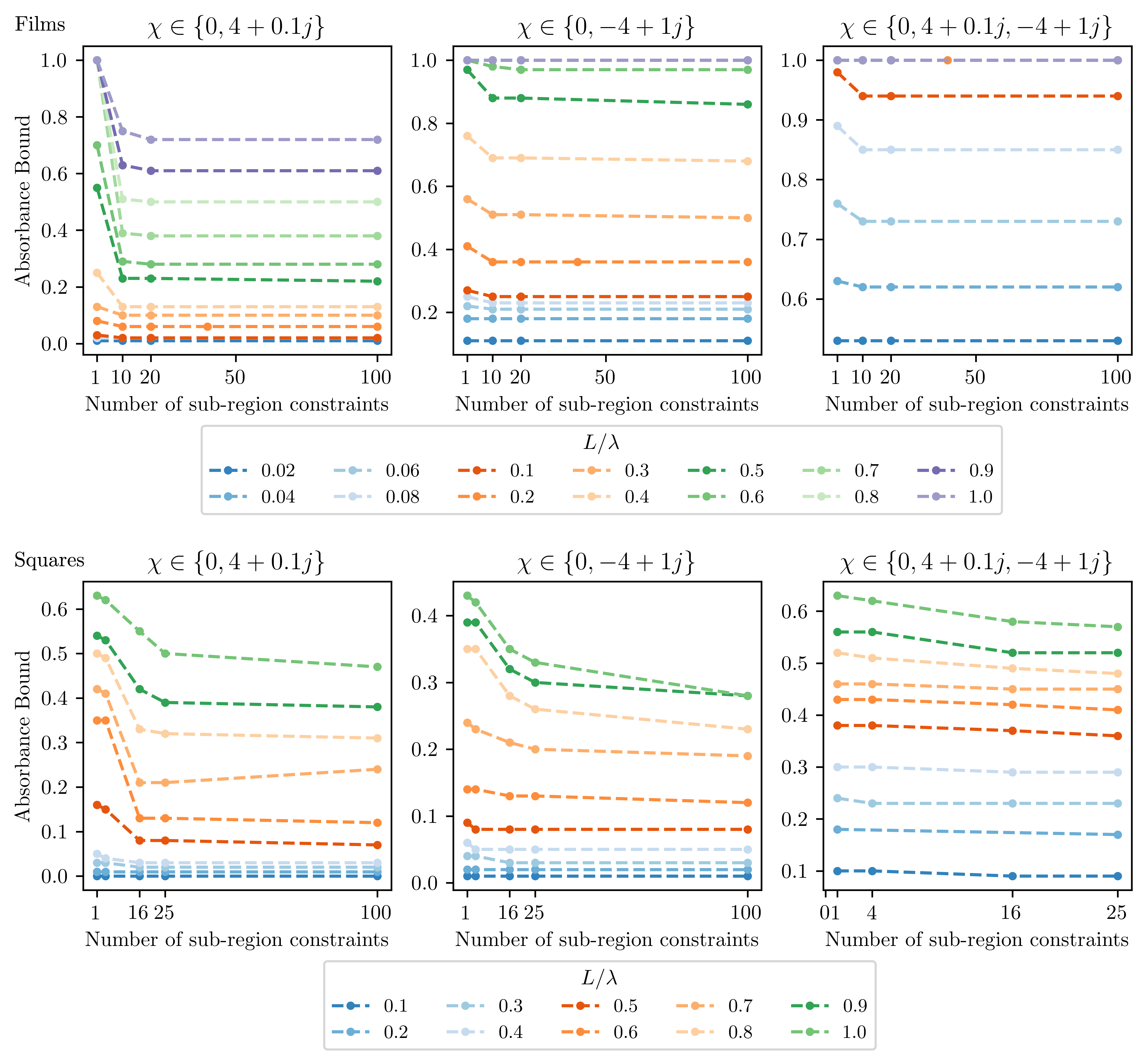}
    \caption{Dependence of calculated absorbance bound on the number of sub-region constraints (i.e. number of projector \( \Proj_j \) constraints enforced in Eq \eqref{eq:optproblem}) for multilayer films (top) and compact structures (bottom). We note that multi-material multilayer films exhibit a perfect absorbance bound for \(L/\lambda \geq 0.2\).}
    \label{fig:constraints}
\end{figure}

In this section, we calculate the dependence of presented bounds on the number of sub-region constraints. For a fixed design size, we can enforce these constraints at finer regions in space through the use of additional projections~\(\Proj_j\). Any additional Lagrange multipliers can be set to \(0\), so enforcing additional constraints exclusively decreases the calculated bound. As a result, increasing the number of sub-region constraints tightens the bound on maximum photonic performance. 

Results are shown in Fig.~\ref{fig:constraints}. We note that many sub-region constraints are not necessary to capture the majority of the physics when \(L/\lambda\) is small. Single material bounds that show imperfect absorbance exhibit a critical number of constraints around which the bound rapidly decreases and saturates. In the case of compact regions (bottom), single material bounds may not have converged for all \(L/\lambda\), indicating that bounds may be possibly made tighter by increasing the number of sub-region constraints. 
Neither the multilayer film nor the compact region heterostructure bounds significantly change over the studied range of number of constraints. This may mean that these bounds cannot be tightened by enforcing local physics, or that the critical point of sub-region constraints has not been reached and bounds could be significantly improved by greater computational efficiency. This remains an open question and will be further studied. 

We note an anomalous point: the compact structure dielectric bound (\(L/\lambda=0.7\)) exhibits a greater bound with \(100\) than with \(25\) sub-region constraints. 
This is a numerical artifact due to the difficulty of optimizing the dual surface, even if convex, over so many parameters and again highlights the need for more efficient numerical methods. 

\section{Inverse Design Details}
\subsection{Multiple Material Topology Optimization}
All calculations were run at increasing resolutions until converged. Multiple material topology optimization was done by writing \(\epsilon = \epsilon_2 + (\epsilon_1 + (\epsilon_{\mathrm{background}} - \epsilon_1)\rho_2 - \epsilon_2)\rho_1\) for \(\rho_1,\rho_2 \in \lbrack 0, 1\rbrack\) and optimizing over the continuous variables \(\rho_1, \rho_2\). The derivatives of the objective with respect to modifications in \(\rho_1, \rho_2\) were computed with Ceviche~\cite{hughes2019forward}. The resulting optimization problem was solved with NLopt \cite{NLopt}.

Inverse designs are often binarized to reflect realistic devices. In the multi-material case, we define a binarized design as one where each pixel is exclusively one of $\epsilon_1$, $\epsilon_2$, or $\epsilon_{\mathrm{background}}$. To better compare with bounds (which at high enough resolutions mimic the behavior of non-binarized devices), inverse designs were not deliberately binarized when calculating device performance. Some designs were binarized for presentation as described in the next section.

\subsection{Inverse Designs in the Main Text}
\label{sec:inverse_design}
\begin{figure}[hbt!]
    \centering
    \includegraphics[width=0.9\linewidth]{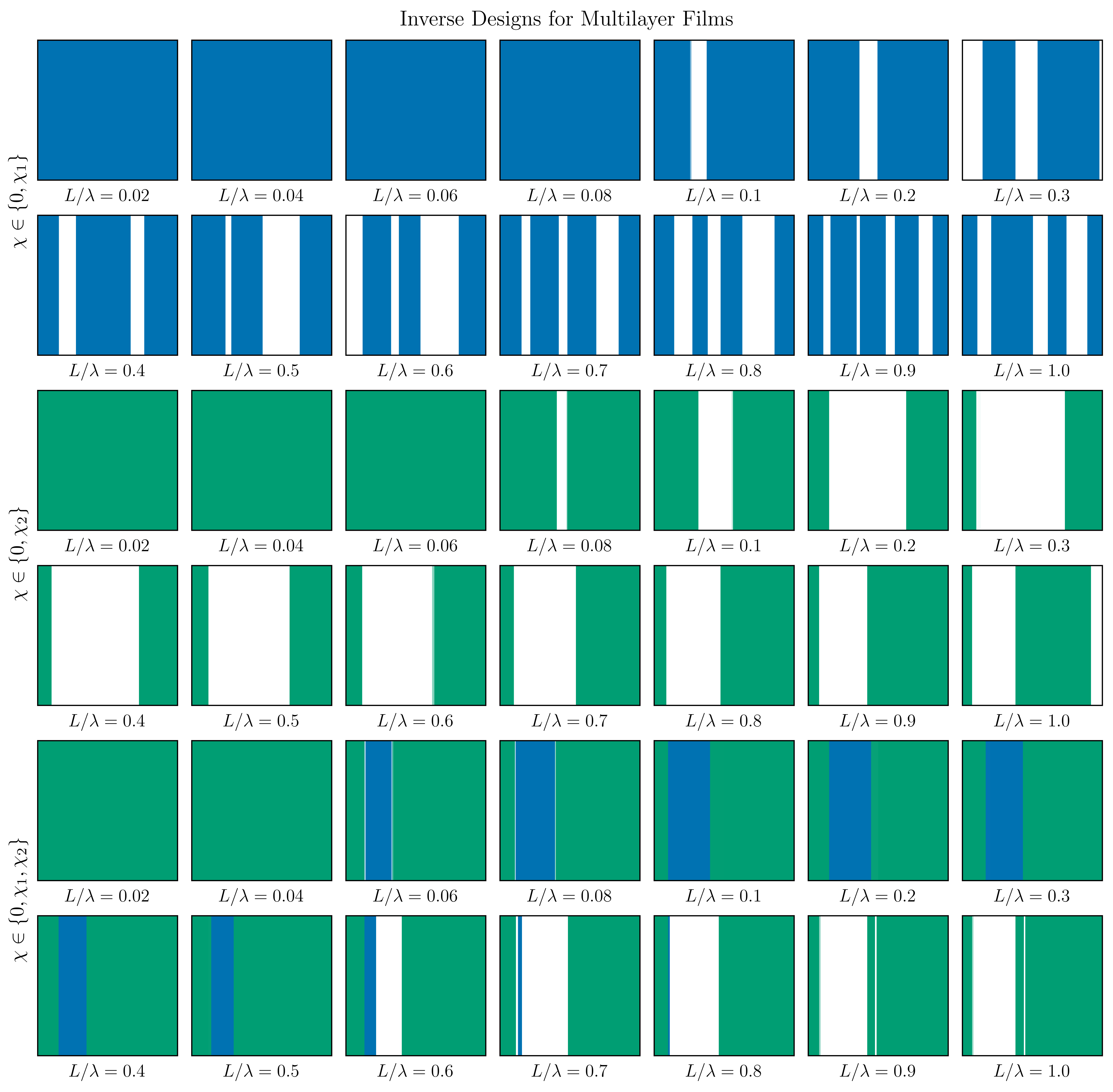}
    \caption{Topology optimized multilayer films comprising materials \( \chi_1 = 4+0.1i \) and/or \(\chi_2=-4+1i \) subject to an incident planewave of wavelength \( \lambda\). Absorbance values are shown in the main text.}
    \label{fig:inverse_designs_films}
\end{figure}

\begin{figure}[hbt!]
    \centering
    \includegraphics[width=0.7\linewidth]{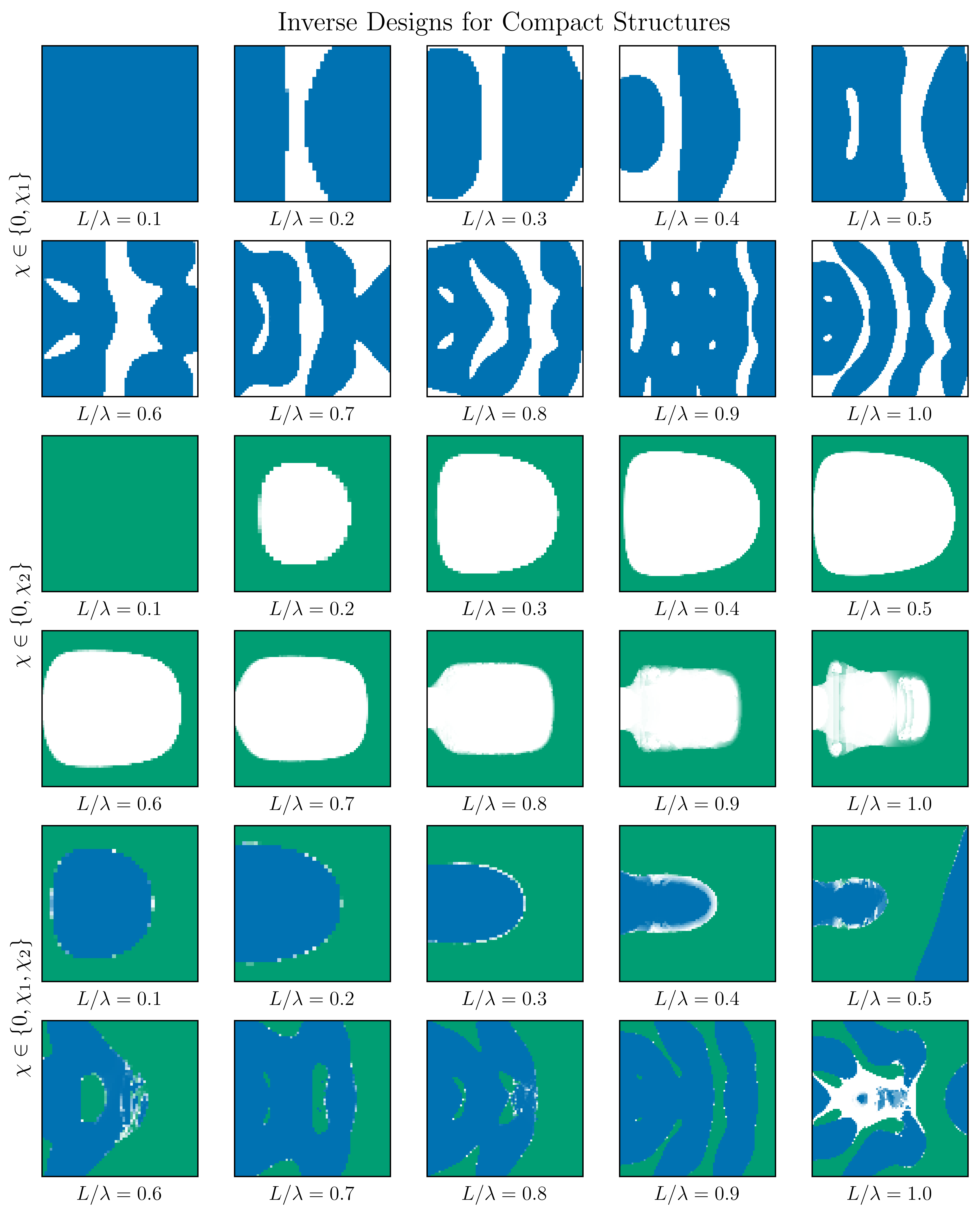}
    \caption{Topology optimized compact structures comprising materials \( \chi_1 = 4+0.1i \) and/or \(\chi_2=-4+1i \) a distance \( d=0.4\lambda \) from an oscillating point dipole with wavelength \( \lambda \). Absorbance values are shown in the main text.}
    \label{fig:inverse_designs_rect}
\end{figure}

All topology optimized inverse designs compared to bounds are shown in Fig. \ref{fig:inverse_designs_films} for multilayer films and Fig. \ref{fig:inverse_designs_rect} for compact structures. 
\textit{Multilayer films:} Designs were binarized for the purposes of presentation if a binarized device could be found with performance within 1\% (metals) and 2\% (multi-material) of the non-binarized device. Devices solely composed of dielectric and vacuum were not deliberately binarized. \textit{Compact structures: } Designs were not binarized.

The heterostructures showcase other examples where vacuum was not utilized, and therefore where these results could have been replicated in a single-material framework with a design region comprised by dielectric or metallic background and surrounded by vacuum. This phenomenon occurred when the single material metallic design vastly outperformed dielectric structures while still exhibiting non-perfect absorbance. These designs are simply metallic cavities filled by an absorbing dielectric. Interestingly, as described in the main text, the dielectric structure outperformed the metallic structure and the corresponding heterostructure utilized vacuum when \(L/\lambda=1\). 

\bibliographystyle{plain}
\bibliography{citations}